\documentclass[12pt,a4paper,final]{iopart}

\usepackage{iopams}  
\usepackage{graphicx}
\usepackage{etoolbox}
\usepackage{upgreek}
\usepackage{braket}
\usepackage[breaklinks=true,colorlinks=true,linkcolor=blue,urlcolor=blue,citecolor=blue]{hyperref}

\usepackage[utf8]{inputenc}
\usepackage[english]{babel}
\usepackage{cite}

\begin{document}

\title[Optical nanofibre-mediated electric quadrupole transition in cold $^{87}$Rb]{Observation of the  $^{87}$Rb $5S_{1/2}$ to $4D_{3/2}$ electric quadrupole transition at 516.6~nm mediated via an optical nanofibre\\}

\author{Tridib Ray$^1$\footnote{present address: Laboratoire Kastler Brossel, Sorbonne Universit\'{e}, CNRS,
ENS-Universit\'{e} PSL, Coll\`{e}ge de France, 4 place Jussieu, F-75005 Paris, France}, Ratnesh K. Gupta$^1$, Vandna Gokhroo$^1$, Jesse L. Everett$^1$, Thomas Nieddu$^1 \textcolor{blue}{ \ddagger}$, Krishnapriya S. Rajasree$^1$, S\'{i}le {Nic Chormaic}$^{1,2}$}
\address{$^1$Light-Matter Interactions for Quantum Technologies Unit, Okinawa Institute of Science and Technology Graduate University, Onna, Okinawa 904-0495, Japan}
\address{$^2$Universit\'e Grenoble Alpes, CNRS, Grenoble INP, Institut N\'eel, 38000 Grenoble, France}
%\address{$^3$School of Chemistry and Physics, University of Kwazulu-Natal, Durban 4000, South Africa}
\ead{sile.nicchormaic@oist.jp}

\begin{abstract}
Light guided by an optical nanofibre can have a  very steep evanescent field gradient extending from the fibre surface.  This can be exploited to drive electric quadrupole transitions in nearby quantum emitters. In this paper, we report on the observation of the $5S_{1/2}$ $\rightarrow$ $4D_{3/2}$ electric quadrupole transition at 516.6~nm (in vacuum) in laser-cooled $^{87}$Rb atoms using only a few $\mu$W of laser power propagating through an optical nanofibre embedded in the atom  cloud. This work extends the range of applications for optical nanofibres in atomic physics to include more fundamental tests.

\end{abstract}

%\pacs{32.30.−r, 42.81.−i, 32.00.00}
\vspace{2pc}

%\submitto{\NJP}
%\maketitle

\section{Introduction}
The simplest and generally the strongest interaction between electromagnetic waves and matter is the electric dipole (E1) interaction - the first radiating term in a multipole expansion.  Naturally, optical excitations  mostly exploit E1 allowed transitions \cite{Demtroder}, such as the cooling transition for alkali atoms. The commonly used selection rules for atomic transitions are based on the electric dipole approximation and the effect of the higher order terms are frequently neglected, leading to what are known as dipole forbidden transitions. The strength of the optical transitions  can be defined in terms of several different parameters, such as the Einstein A and B coefficients, the dipole moments, or the  oscillator strengths (i.e. the $f$ values) \cite{doi:10.1119/1.12937}. The dipole Rabi frequency, which defines the actual number of transitions that take place per second in the two-level system, is linearly proportional to the dipole moment and the amplitude of the light field.  The corresponding E1 oscillator strengths (i.e. the $f$ values) are proportional to the square of the dipole moment and, hence, the square of the E1 Rabi frequency.  

The next term in the multipole expansion is the electric quadrupole (E2).  Quadrupole transitions play an important role in atomic and molecular spectroscopy \cite{sayer_absorption_1971,camy-peyret_quadrupole_1981,tojo_absorption_2004} with relevance in photochemistry, atmospheric physics, and fundamental processes \cite{PhysRevA.89.012502}, to name just a few. The E2 Rabi frequency is linearly proportional to the quadrupole moment and the gradient of the light field. Similarly  to the E1 case, the oscillator strength for E2 transitions is proportional to the square of the E2 Rabi frequency. This implies that the quadrupole oscillator strengths also depend on the gradient of the electric field. Due to this dependency, E2 transitions are less studied  than their  E1 counterparts as it can be challenging to create a large enough field gradient experimentally.   Note that the square of the quadrupole Rabi frequency is equivalently proportional to the E2 oscillator strength.  

Several platforms for driving E2 transitions in the alkali atoms have been proposed and/or demonstrated, with recent particular focus on the $S \rightarrow D$ transitions since they may be useful for high-precision measurements of parity non-conservation (PNC) \cite{PhysRevA.89.012502} and could be used for an exchange of orbital angular momentum between light and the internal states of the atom \cite{PhysRevLett.89.143601, le_kien_enhancement_2018}.  There are significant studies on the $6S_{1/2}$ to $5D_{5/2}$ transition in Cs at 685~nm using a variety of techniques including  evanescent light fields from prism surfaces \cite{tojo_absorption_2004,tojo_precision_2005}, surface plasmons\cite{chan_coupling_2019}, and continuous wave (CW) free-space excitation \cite{pucher2019precise}, with proposals for using optical vortices \cite{lembessis_enhanced_2013}, nano-edges \cite{shibata_excitation_2017}, and plasmonics \cite{sakai, chan_coupling_2019}.  

Experiments in rubidium have been more limited, with pulsed laser excitation of the $5S_{1/2}$ to the  $4D$ levels at 516.5~nm \cite{1978A&A....64..215T, NILSEN1978327} and from the $5S_{1/2}$ to $nD$, where  $n=27-59$ Rydberg levels using pulsed excitation at $\sim$ 297~nm \cite{PhysRevA.79.052509} being reported.  Some works have also focussed on $n_1^2P \rightarrow n_2^2P$ electric quadrupole transitions in Rb  by exploiting double resonances \cite{PhysRevA.92.042511, Mojica_Casique_2016}.  However, for the reasons mentioned earlier,  our focus is primarily on the $S \rightarrow D$ quadrupole transitions.  The difficulty in exciting the $5S_{1/2} \rightarrow 4D$ transitions lies in the fact that the ratio of E1 to E2 transitions in the visible region is of the order of $10^7$, so aside from the desirability of a strong electric field gradient, sufficient laser intensity at $\sim$ 516.6~nm is needed to drive the transition, with the cross-section for absorbing one photon being $1.4 \times 10^{-17}$ cm$^2$ \cite{1978A&A....64..215T}.     

In this paper, we report on the experimental observation of the 516.6~nm (in vacuum) electric quadrupole transition in a cloud of laser-cooled $^{87}$Rb atoms mediated by an optical nanofibre (ONF) using CW light. The rapid radial exponential decay of the 516.6~nm evanescent field from the surface of the ONF provides a very steep electric field gradient in the region of highest field intensity even for very low excitation laser powers, leading to relatively efficient excitation of the E2 transition.  Additionally, optical nanofibres are relatively easy to fabricate and integrate into magneto-optical traps or atomic vapour cells, as evidenced by the sheer volume of work in the last decade  \cite{ hendrickson_observation_2010, vetsch_optical_2010,nayak_spectroscopy_2012, lacroute_state-insensitive_2012,   kumar_multi-level_2015, kumar_autler-townes_2015, sayrin_storage_2015, kumar_temperature_2016, ruddell_collective_2017, beguin_observation_2018, kato_observation_2019,  rajasree_generation_2019}, negating the necessity for nanofabrication facilities as needed when using metamaterials or other such nanostructures.

\section{Theoretical Considerations}

\begin{figure}
    \centering
    \includegraphics[width=0.42\linewidth]{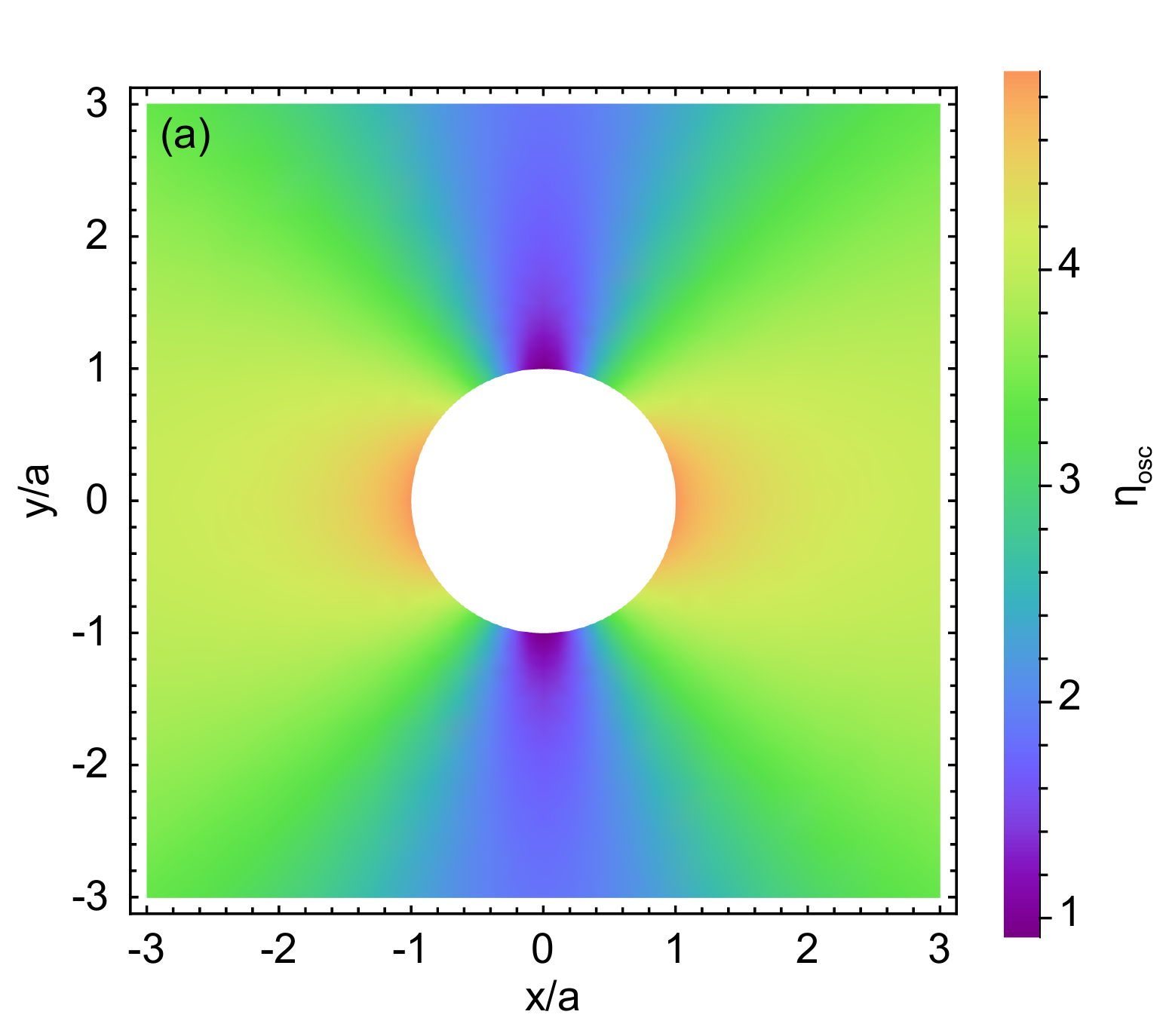}\includegraphics[width=0.58\linewidth]{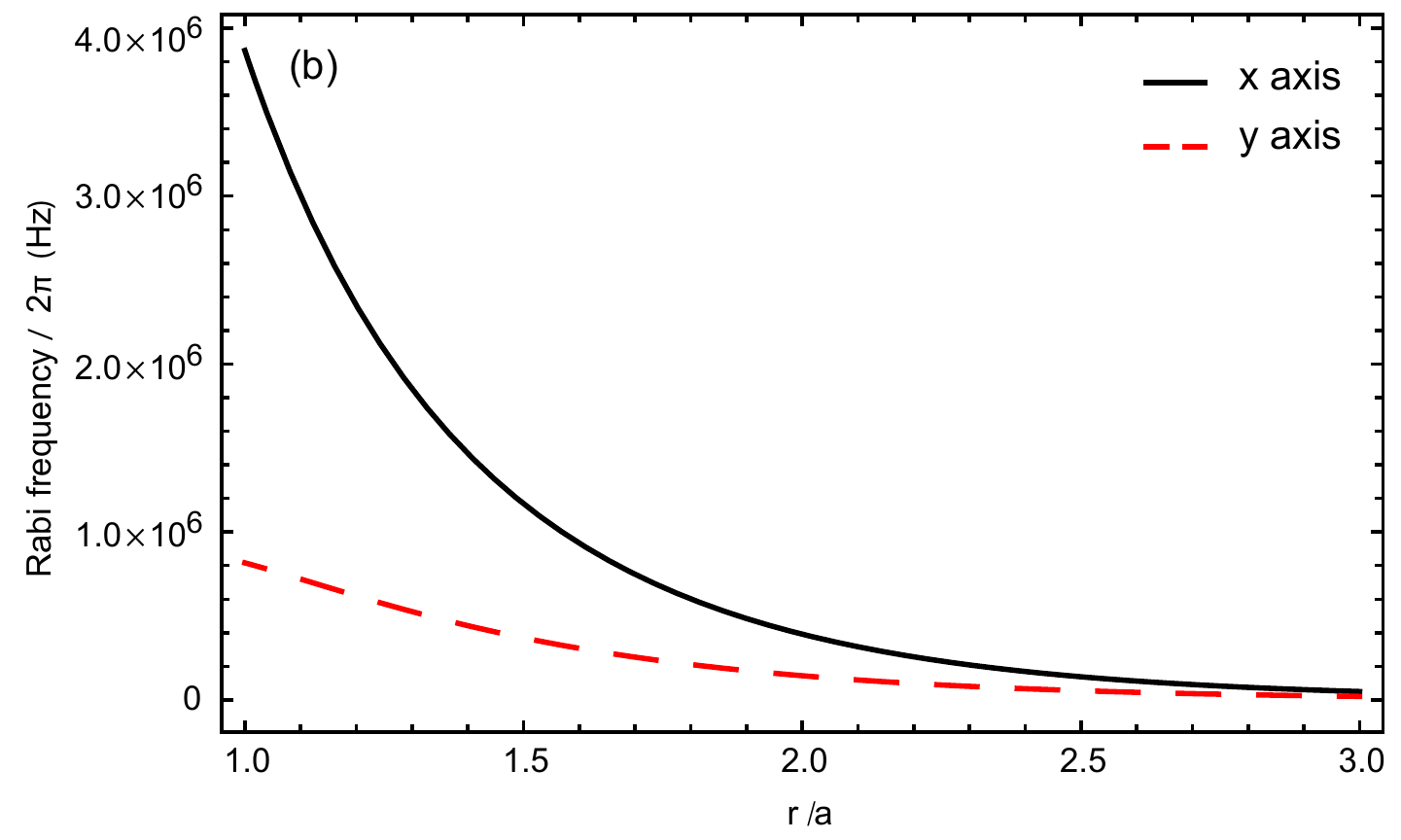}
    \caption{(a) The quadrupole oscillator strength enhancement factor $\eta_{osc}$ for a 200 nm radius optical nanofibre, with quasilinear polarisation along the $x$-axis. (b) Quadrupole Rabi frequency for the $5S_{1/2}(F=2) \rightarrow 4D_{3/2}(F''=3)$ transition with 1 $\mu$W of light at 516.6~nm propagating in the fibre, for atoms positioned along the $x$-axis (solid line) or $y$-axis (dashed line).}
    \label{fig:theory}
\end{figure}

Recently, we  theoretically investigated the enhancement of the $5S_{1/2} \rightarrow 4D_{5/2}$ quadrupole interaction for a $^{87}$Rb atom in the evanescent field of an optical nanofibre \cite{le_kien_enhancement_2018}. We proposed that, while the E2 Rabi frequency reduces rapidly with radial distance from the fibre, the E2 oscillator strength enhancement is still significant even at appreciable distances  from the nanofibre surface. In Ref. \cite{le_kien_enhancement_2018}, the oscillator strength enhancement was expressed in terms of an \emph{enhancement factor}, $\eta_{osc}$, which is the ratio of the oscillator strength of a fibre-guided field to the oscillator strength defined for a free-space, plane wave field with equal intensity. The enhancement depends on the wavelength and the fibre geometry, not the transition itself.   Following the methods from Ref. \cite{le_kien_enhancement_2018}, here we consider a different E2 excitation in $^{87}$Rb, namely the $5S_{1/2} \rightarrow 4D_{3/2}$ transition.  We focus on this transition due to technical issues related to photon detection for the subsequent decay channels.   In Fig. \ref{fig:theory}(a), $\eta_{osc}$ is plotted as a function of atom position in the $xy$-plane (where we assume the fibre axis is along $z$ and the axis for the quasilinearly polarised guided light is along $x$).  Assuming  light is guided in the fundamental fibre mode, ${H\!E}_{11}$, we find a maximum enhancement factor for the oscillator strength, $\eta_{osc} = 4.92$,  for an atom located on the fibre surface and positioned along the $x$-axis.  As the atom moves further from the fibre, the enhancement still exists as long it is positioned close to the  $x$-axis.  This enhancement factor is not easily measured experimentally since it is a comparison between the ONF-mediated quadrupole transition and that in free-space for a specific intensity at a single point; it does not account for the exponential decay of the evanescent field.  

If we take the varying profile of the evanescent field into consideration, a far more experimentally useful and accessible parameter is the quadrupole Rabi frequency, which, for simplicity, we refer to as the Rabi frequency in the following. For an atom with ground state, $g$, and excited state, $e$, the Rabi frequency is given by \cite{le_kien_enhancement_2018} \begin{equation}
    \Omega_{ge}=\frac{1}{6\hbar}\sum_{ij}{\bra{e}\mathcal{Q}_{ij}\ket{g}\frac{\partial \mathcal{E}_j}{\partial x_i}},
\end{equation}
where the $\mathcal{Q}_{ij}$ are the quadrupole tensor components representing the strength of the quadrupole transition.  In Fig. \ref{fig:theory}(b), we plot the Rabi frequency as a function of an atom's position relative to the fibre, either along or perpendicular to the quasi-polarisation axis.  We choose 1 $\mu$W of resonant optical power (at 516.6~nm) transmitted in the fundamental mode of the nanofibre. We see that the Rabi frequency reduces dramatically as the atom moves away from the fibre surface.

\section{Experiments}

\begin{figure*}[t]
\center
\includegraphics[width=16 cm]{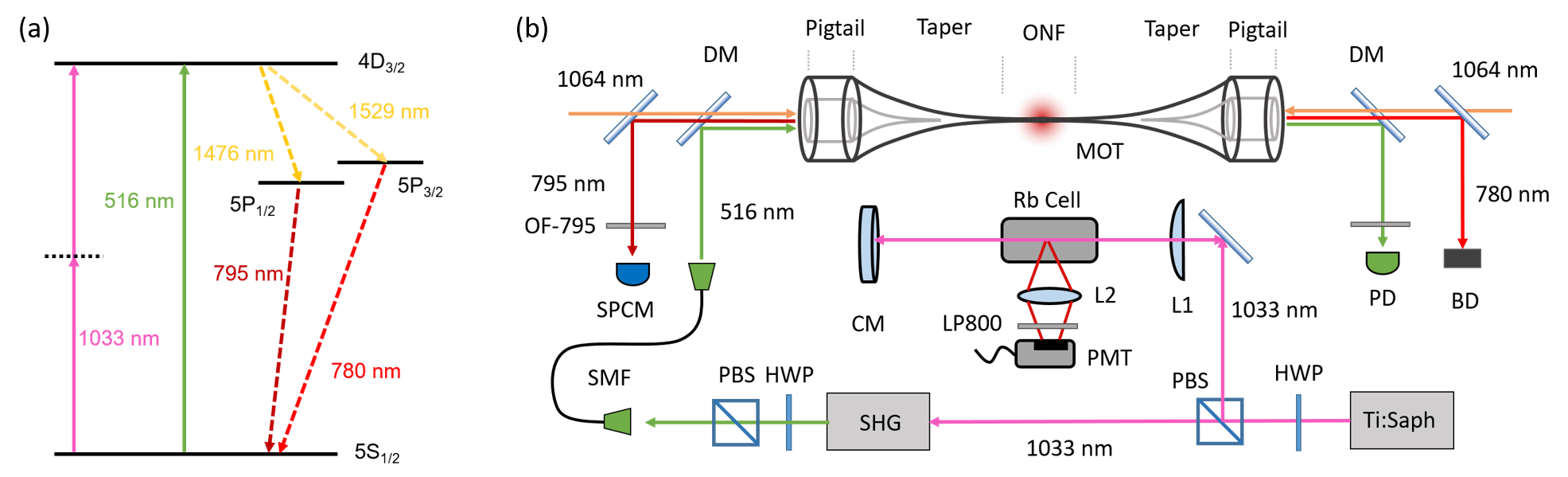}
\caption{(a) Simplified energy level diagram for $^{87}$Rb .  The quadrupole transition from $5S_{1/2} \rightarrow 4D_{3/2} $ can be driven using a single photon at 516.6~nm or two photons at 1033.3~nm.  (b) Schematic experimental setup.  ONF: Optical nanofibre, PBS: Polarising beamsplitter, HWP: Half-wave plate, SHG: Second harmonic generator, PD: Photodiode, BD: Beam-dump, DM: Dichroic mirror, SMF: Single-mode fibre, SPCM: Single photon counting module, PMT: Photomultiplier tube, LP800: Longpass filter}
\label{fig:schematic}
\end{figure*}

In our experiment, we  drive the $5S_{1/2} \rightarrow 4D_{3/2}$ E1 forbidden, E2 allowed optical transition in a cold atomic ensemble of $^{87}$Rb, via the evanescent field of an optical nanofibre embedded in the atom cloud. The relevant energy level diagram is shown in Fig. \ref{fig:schematic}(a) and a schematic of the experiment is given in Fig. \ref{fig:schematic}(b). The E1 forbidden transition is excited using a single photon transition at 516.6~nm derived from a frequency-doubled Ti:Sapphire laser (M Squared SolsTiS laser and ECD-X second harmonic generator) set to 1033.3~nm.  The 516.6~nm light from the second harmonic generator (SHG) is mode cleaned using a single-mode fibre and is coupled to one pigtail of the ONF using a pair of dichroic mirrors, see Fig. \ref{fig:schematic}(b). Shortpass filters are placed after the SHG to remove any residual 1033.3~nm, which could lead to two-photon excitation of the transition of interest.   We  control the 516.6~nm power through the ONF using a half-wave plate combined with a  polarisation beam splitter (PBS) and neutral density filters (NDF). 

Atoms excited to the $4D_{3/2}$ state can decay back to the ground state via two channels, see Fig. \ref{fig:schematic}(a): (i) the $5P_{1/2}$ intermediate state, by cascaded emission of two photons at 1476 nm and 795 nm and (ii) the $5P_{3/2}$ intermediate state, by cascaded emission of two photons at 1529~nm and 780~nm \cite{Moon:11, Roy:17}, which couple into the ONF and can be detected at the output pigtail. Detection of either of the emitted photon pairs would allow us to infer the electric quadrupole excitation. In this work, we detect the second step of decay path (i) at 795~nm due to the  availability of single photon detectors (SPD) at  near infrared (NIR) wavelengths and its spectral separability from the 780 nm photons scattered during the atom cooling process.  The first step in the decay path at  1476~nm is undetected.  

\begin{figure}[ht]
\center
\includegraphics[width=8 cm]{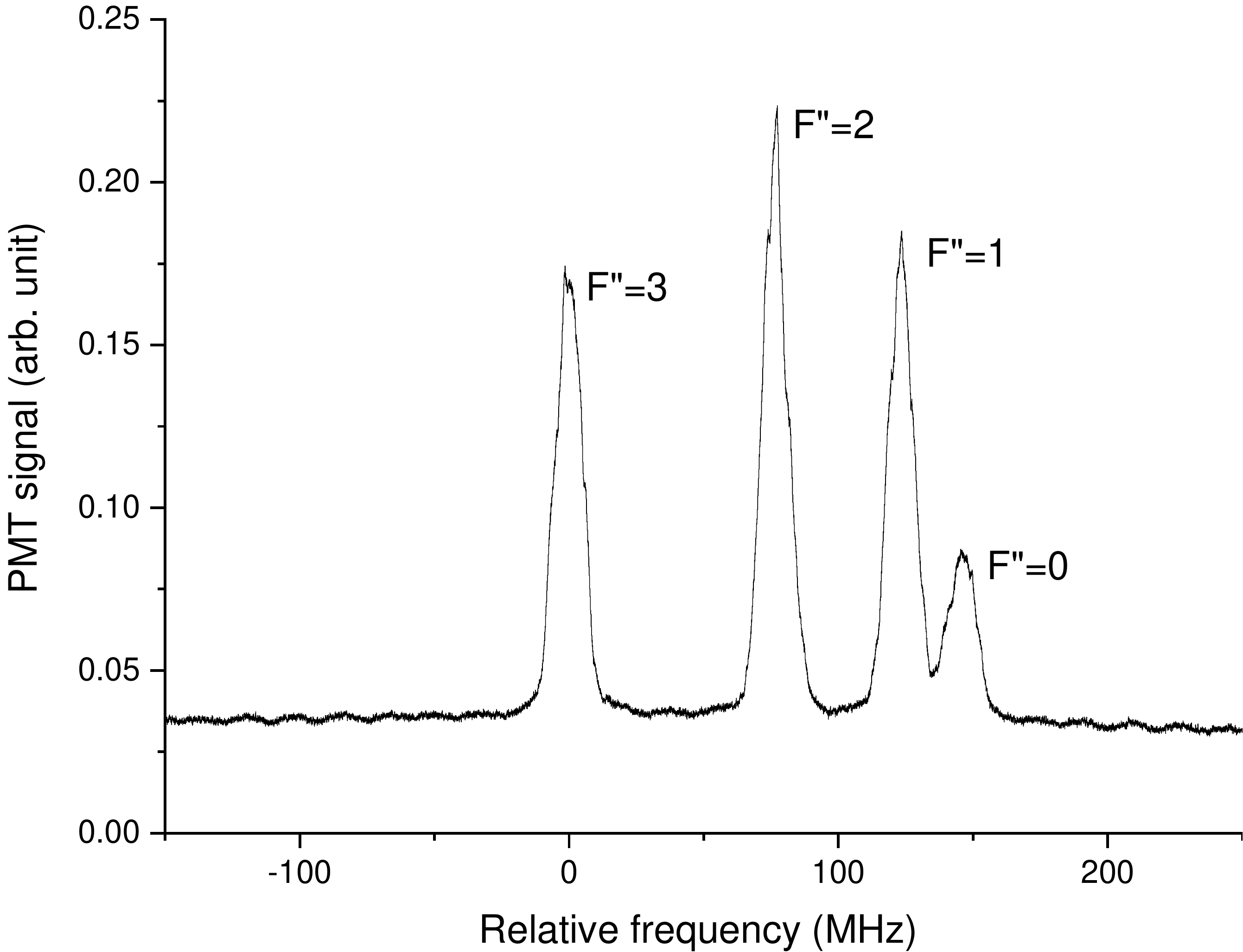}
\caption{Vapour cell spectroscopy for the $5S_{1/2}\ F=2$ to the excited states $4D_{3/2}\ F''$ using free-space two-photon excitation at 1033.3~nm.}
\label{vapor cell spectroscopy}
\end{figure}

A frequency reference for the quadrupole transition is obtained using free-space, two-photon spectroscopy in a Rb vapour cell heated to 125$^{\circ}$C \cite{Nieddu:19}. We use the tightly focussed, 1033.3~nm direct output from the Ti-sapphire laser to drive the two-photon transition.  The laser is scanned around 1033.3~nm such that atoms are excited to the $4D_{3/2} F''$ level. Figure \ref{vapor cell spectroscopy}  shows the two-photon spectroscopy signals for $5S_{1/2} F=2$ to $4D_{3/2} F''$. 
The four observed peaks correspond to the allowed transitions  for the two-photon excitation at 1033.3~nm ($\Delta F 	\leq 2)$ \cite{SALOUR1978364}.   

To study the quadrupole transition itself, we use $^{87}$Rb atoms that are cooled and trapped in a conventional 3D magneto-optical trap (MOT) to a temperature of about $\sim$ 120 $\mu$K.  The cold atom cloud has a Gaussian full width at half maximum (FWHM) of $\sim$  0.5~mm.  We fabricate the ONF, which has a waist diameter of $\sim$ 400~nm, by exponentially tapering a section of SM800-125 fibre using the flame-brushing method \cite{doi:10.1063/1.4901098}.   Further details of the MOT setup and the ONF can be found elsewhere \cite{rajasree_generation_2019}.  The ONF has been optimised for 780~nm propagation and only  $\sim 10\%$ of 516.6~nm light (used to drive the E2 transition) coupled into the ONF pigtails is detected at the output.  We assume that most losses are at the down-taper as the adiabatic criterion is not satisfied for this wavelength.  For wavelengths shorter than $\sim$ 550~nm, the ONF  supports several guided modes. Hence, for light at 516.6~nm, the ONF can support the HE$_{11}$ mode and the first group of higher order modes, TE$_{01}$, TM$_{01}$, and HE$_{21,eo}$  \cite{FRAWLEY20124648}.  We have seen elsewhere that, by coupling the HE$_{11}$ mode into the input fibre pigtail to the ONF, the amount of coupling into the higher order modes is minimal \cite{Nieddu2019, Mekhail2019} and we, therefore, assume that the atoms only interact with the fundamental mode at the nanofibre waist.  Note that, in all the following, when we refer to light power through the ONF, the values given are those measured at the output pigtail of the fibre and not at the nanofibre waist.  This may be higher, due to both the aforementioned design criteria and, additionally, wavelength-specific material losses.   

In addition to the fibre-guided light at 516.6~nm, we also inject a pair of 1064~nm counter-propagating beams into the fibre.  This has several purposes.  First, it keeps the ONF hot during the experiments, thus avoiding atom deposition on the fibre.  Second, it  attracts atoms towards the ONF surface due to the optical dipole force, thereby increasing the number of atoms in the evanescent field, resulting in larger photon signals.

\section{Results and Discussion}

For a demonstration of the quadrupole excitation in the cold atom-ONF system,   we study the $4D_{3/2} F''$ transition in detail.  As mentioned, atoms can decay from  $4D_{3/2}F''$ to $ 5S_{1/2} F$ via $4P_{1/2}$  emitting  795 nm photons. These emitted photons can be filtered efficiently from the 780 nm photons scattered from the trapping light leading to a relatively clean signal. 
The experiment is performed by scanning the Ti:Sapphire laser frequency across the $4D_{3/2} F''$ transitions, see Fig.~\ref{fig:schematic}, and recording the 795~nm decay photons generated from the ONF-mediated E2 transition on a single-photon counting module (SPCM).  Simultaneously, we record the 1033.3~nm, two-photon spectroscopy signal from the vapour cell using a photomultiplier tube (PMT), see Fig.~\ref{fig:schematic}.   Each experiment cycle is 10 seconds long, with 20~ms of bin time both for the SPCM and the PMT. Each data point is an average of 50 cycles. Note that we can also detect the transmission of the 516.6~nm light through the fibre using a photodiode (PD) if desired. 

\begin{figure}[ht]
\center
\includegraphics[width=8cm]{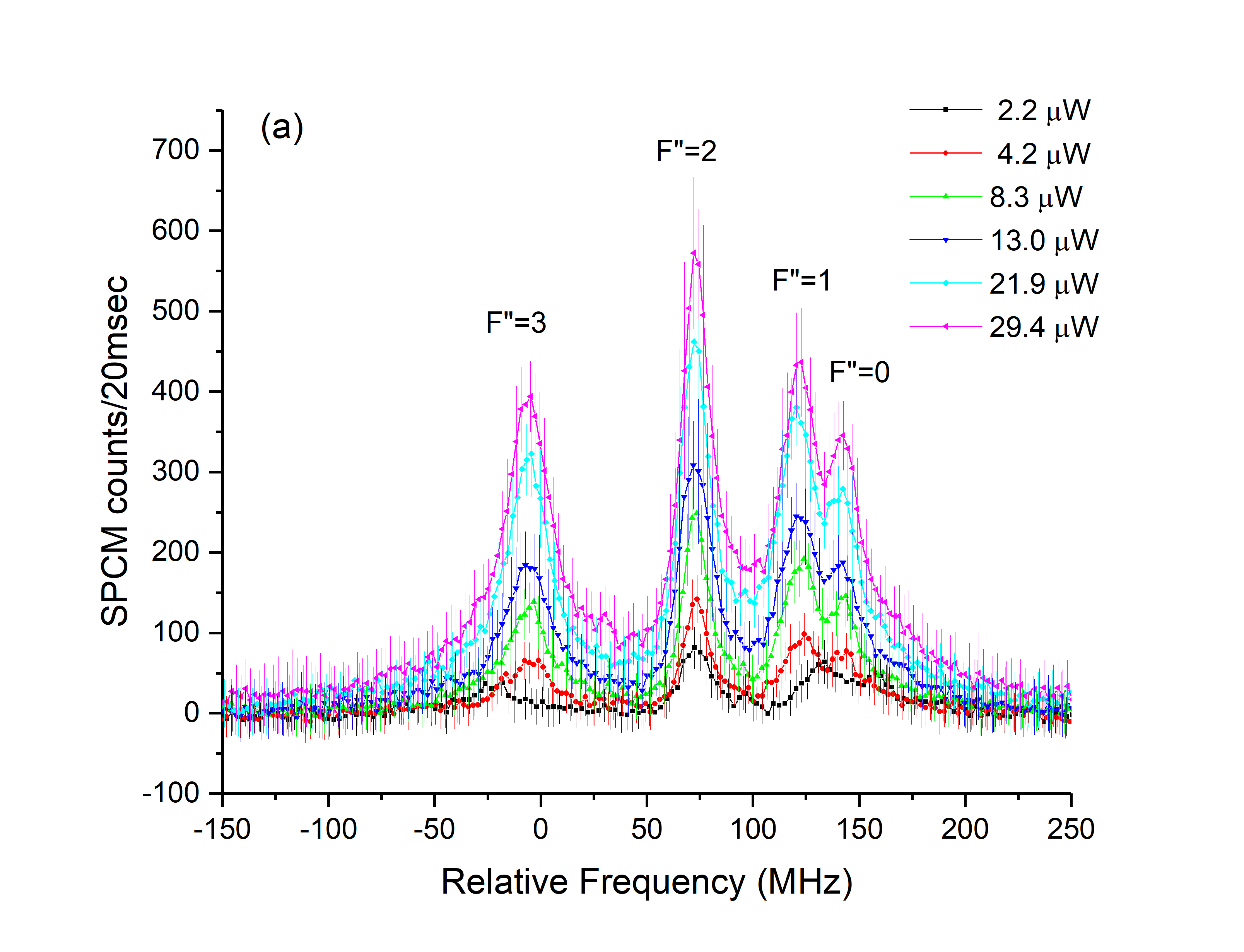}\includegraphics[width=8cm]{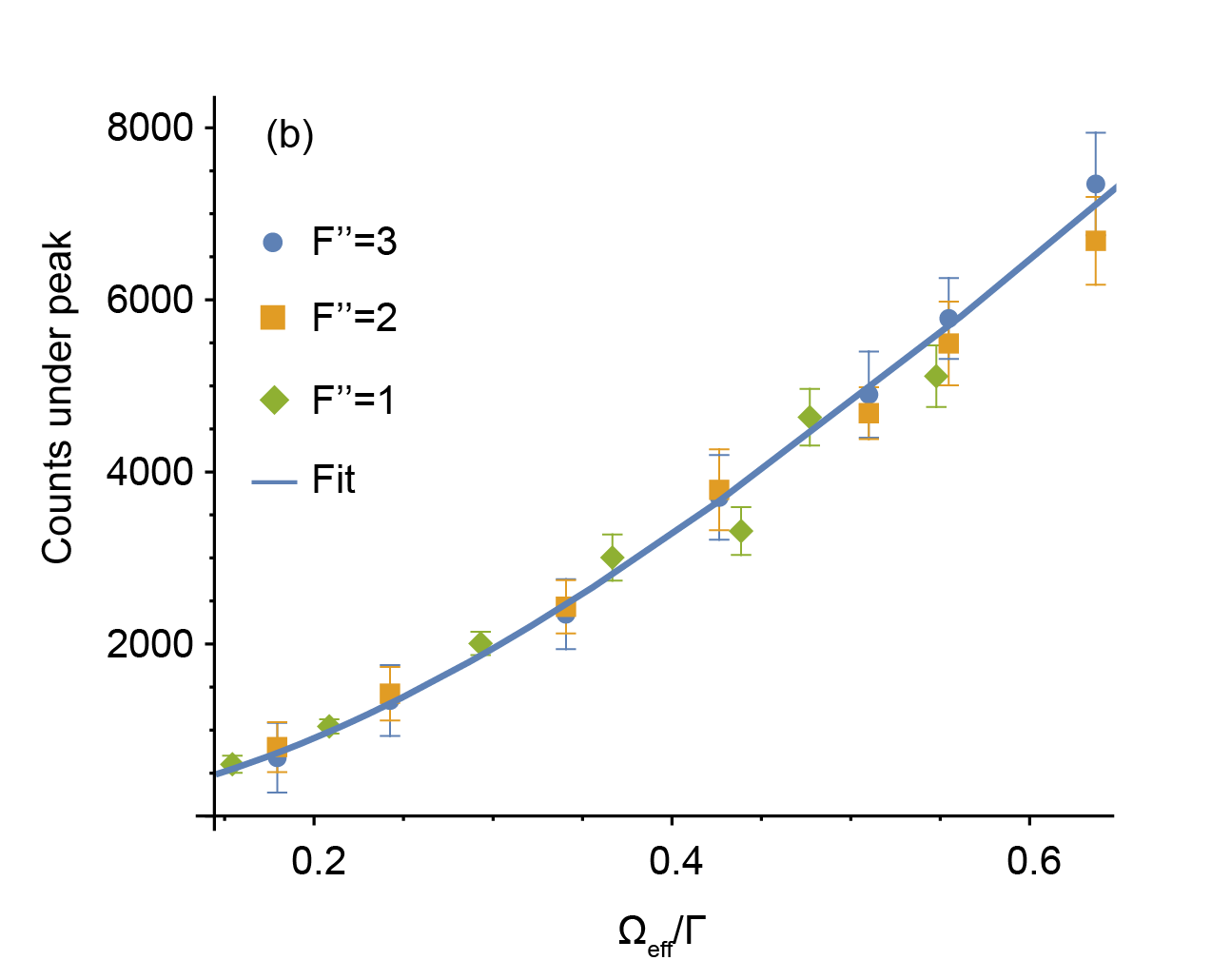}
\caption{Quadrupole transition spectroscopy signal in $^{87}$Rb for the  $5S_{1/2}$ ground state to the $4D_{3/2}$ excited states using 516.6~nm light through the ONF.  (a) The 795~nm signal measured at the SPCM as a function of the 516.6~nm frequency for different powers in the ONF; (b) Photon count under the peak for the first three transitions, plotted against propagating power scaled to the fitted value of $\Omega_{eff}$. Power is measured at the output pigtail of the ONF.}
\label{fig:powerstudy}
\end{figure}

As a first experiment, we studied the dependence of the 795~nm emission on the 516.6~nm power propagating through the ONF, see Fig. \ref{fig:powerstudy}(a).  The four observed peaks correspond to the electric quadrupole transitions, $\Delta F \leq 2$, and they are  comparable to the observed two-photon dipole spectroscopy signal at 1033.3~nm in Fig. \ref{vapor cell spectroscopy}.  Expected features, such as peak broadening and peak shifts  due to the presence of the ONF, are visible in the spectra, with wide asymmetric tails which we attribute to the van der Waals interaction between the nanofibre and the atoms \cite{Minogin2010}. Due to the roughly exponential decay profile of the evanescent field, atoms experience a varying 516.6~nm intensity as they are excited to the $4D_{3/2}$ state. We ignore this and assume that the 795~nm emission into the  fibre is produced by stationary atoms in a constant field with an effective quadrupole Rabi frequency $\Omega_{eff}$. This frequency includes the oscillator strength $f_{FF''}$ for each $F\rightarrow F''$ transition, so each $F''$ level experiences a different $\Omega_{eff}$ for a specific propagating power in the fibre.  

From Fig. \ref{fig:powerstudy}(a), we extract $\Omega_{eff}$ for each power by modelling each $F''$ transition as a broadened Lorentzian. We integrate the photon counts for each transition, with bounds set manually. The $F''=0$ transition is discarded due to the large overlap that it has with $F''=1$. Using the area under the peak allows us to ignore the exact source of the broadening. The integrated photon count is then related to the effective Rabi frequency by $A\propto \Omega_{eff}^2/\sqrt{\Gamma^2+2\Omega_{eff}^2}$, where $\Gamma/2\pi$ is the decay rate of the $4D_{3/2}$ state indirectly towards the ground state. In Fig. \ref{fig:powerstudy}(b) we fit the data to find $\Omega_{eff}=(0.12 \pm0.02)\Gamma$ for 1~$\mu$W of propagating power, with the data plotted directly against the fitted value. Ignoring intermediate state lifetimes and the effect of the nanofibre on E1 transition rates, the dipole decay from the $4D_{3/2}$ state gives $\Gamma/2\pi=2.12$ MHz, resulting in $\Omega_{eff}/2\pi=250\pm50$ kHz for a power, $P$, of 1 $\mu$W in the fibre. For comparison, the theoretically estimated value  for the Rabi frequency  (see Fig. \ref{fig:theory}(b)), averaged over the azimuthal angle, is about 250~kHz for 1~$\mu$W of propagating optical power, 200~nm from the fibre. Since  there is good qualitative agreement between the value predicted by theory and that measured experimentally by our alternate method, we can conclude that the 516.6~nm power at the waist should correspond roughly to that measured at the output.  

\begin{figure}[ht]
\center
\includegraphics[width=8.6cm]{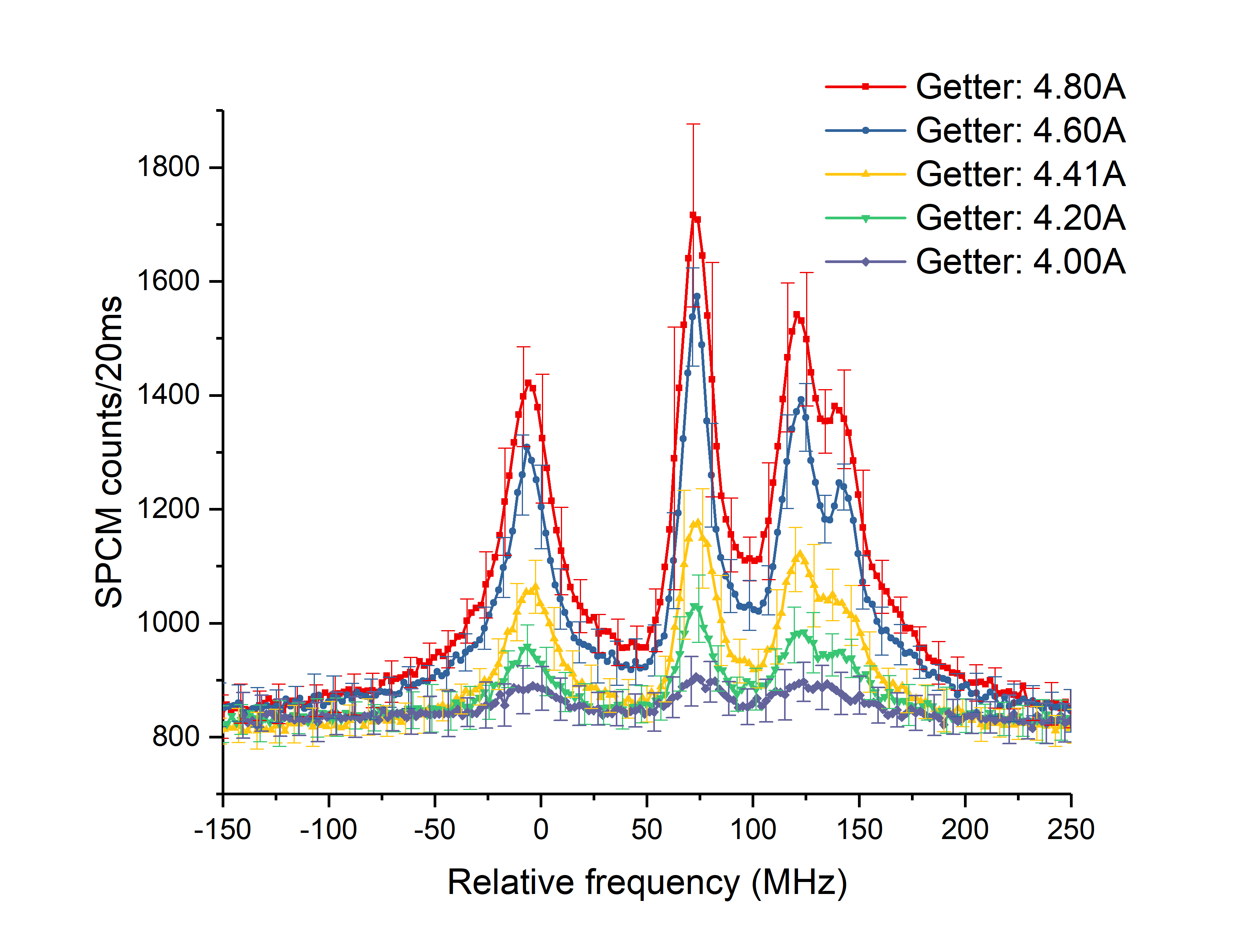}
\caption{Quadrupole transition spectroscopy signal in  $^{87}$Rb for the $5S_{1/2}$ ground state to the  $4D_{3/2}$ excited states using 516.6~nm light through the ONF showing the 795~nm signal as a function of the 516.6~nm frequency for different Rb getter currents. The peaks are the same as in Fig. \ref{fig:powerstudy}.}
\label{fig:getter current}
\end{figure}

In a second experiment, we studied the dependence of the 795~nm emission on the getter current to the Rb source.  Results are shown in Fig. \ref{fig:getter current}.  By assuming a Gaussian profile atom cloud around the ONF, we estimate that a getter current of 4.0 A (4.8 A) corresponds to a density of $\sim 6\times 10^9$ ($8\times 10^9$) atoms/cm$^3$.  Note, however, that the cloud shape changes dramatically during these measurements and it is more accurate to consider the number of atoms in the trap increasing with getter current.   In addition, the size of the atom cloud is increasing along the length of the nanofibre, leading to stronger photon signals.  

\begin{figure}[ht]
\center
\includegraphics[width=8.6cm]{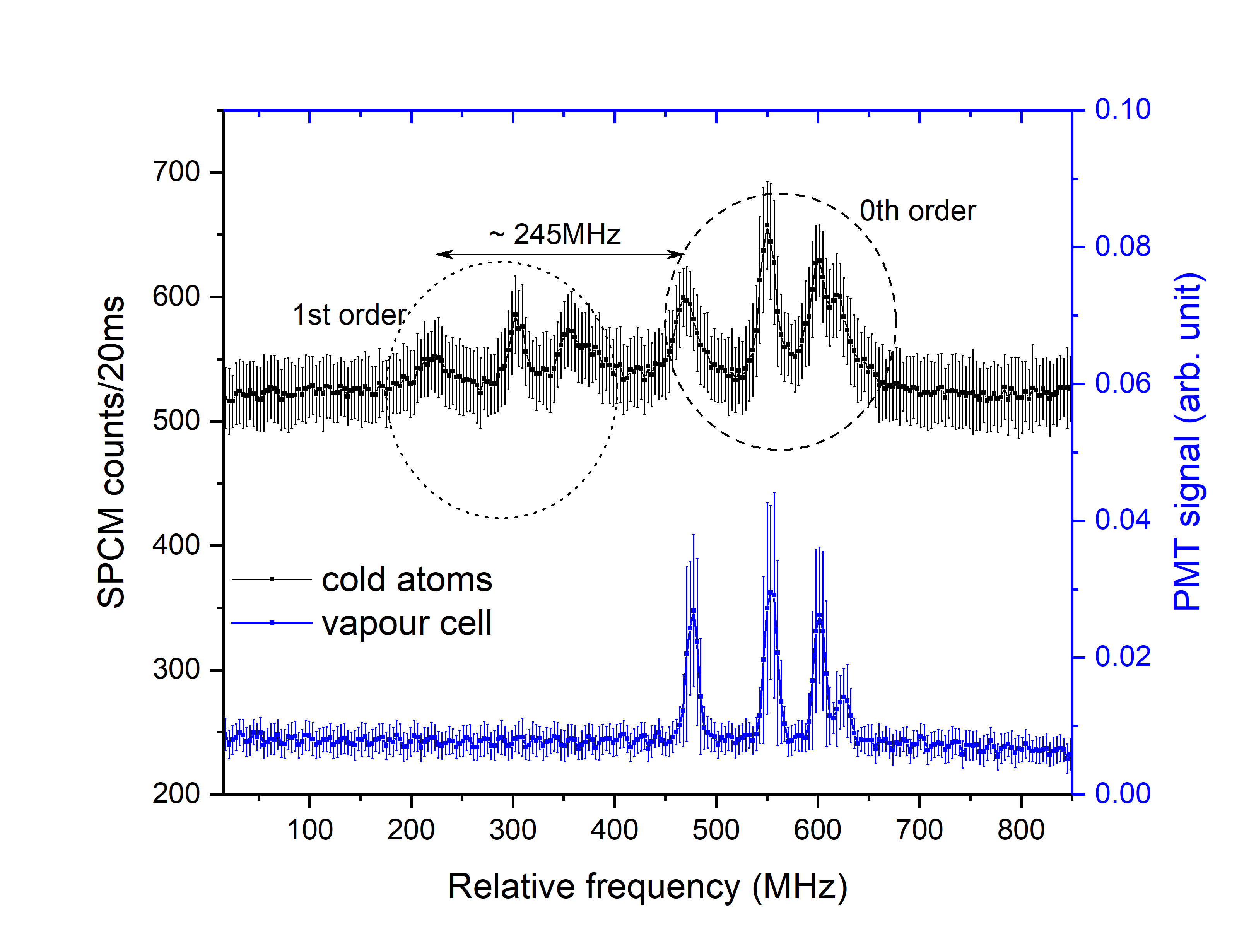}
\caption{ Spectroscopy signal for $5S_{1/2}$ to   $4D_{3/2}$ in $^{87}$Rb.  Bottom: Two-photon reference at 1033.3~nm in a vapour cell.  Top:  Quadrupole excitation in cold atoms when the SHG output is passed through an AOM before being coupled to the ONF.  The presence of the 1st order signal due to the AOM frequency shift, in the absence of a 2nd order signal, verifies the E2 excitation.  The strong 0th order signal is due to inefficient coupling of light into the 1st order.}
\label{aom_test}
\end{figure}

 As a verification that the transition is E2 driven, we used an acousto-optic modulator (AOM), with a central frequency $\omega_{RF}$ $\sim$ 245 MHz, to shift the frequency of the SHG output light.  Here, light from the SHG at 516.6~nm and any residual light at 1033.3~nm passed through the AOM and the positive first-order was sent through the ONF.  The frequency of the light with respect to the SHG output is shifted by $\omega_{RF}$. If the atoms were excited via the E2 transition at 516.6~nm, the spectroscopy signal peaks should shift by $\omega_{RF}$, whereas if excitation was via the two-photon transition by residual light at 1033.3~nm, the peaks should shift by $2 \times \omega_{RF}$.   Observations show, see Fig. \ref{aom_test}, that the signal is shifted by $\omega_{RF}$; thence, the excitation is indeed via the quadrupole transition.

\section{Conclusion}
We have demonstrated an optical nanofibre-mediated electric quadrupole transition, $5S_{1/2}$ to the $4D_{3/2}$, in $^{87}$Rb at 516.6~nm \cite{safronova_critically_2011}, by recording fluorescence emissions at 795~nm.  An important feature is that only a few $\mu$W of power were needed to drive the E2 transition.  Even though the 1476~nm  photon from the first step in the decay path is undetected in this work, it may be possible to indirectly determine the lifetime of the $4D$ level from the fluorescence distribution of the 795~nm step \cite{PhysRevA.72.012502} since the lifetime of the $5P_{1/2}$ level is well-documented \cite{PhysRevA.57.2448, PhysRevA.66.024502}.    This will be the focus of future work.   

The $5S_{1/2}\rightarrow 4D_{X/2}$ transition in $^{87}$Rb could be used to study parity-violating nuclear forces beyond the standard model with the accuracy in Rb expected to be higher than that for Cs \cite{PhysRevA.86.062512, PhysRevA.89.012502} or could be exploited for the transfer of orbital angular momentum of light to the internal degrees of freedom of an atom. A similar technique to the work presented here could be used to study the $4D_{5/2}$ transition; this decays to $5S_{1/2}$ along a single path via the $5P_{3/2}$ state with the emission of 1529~nm and 780~nm correlated photons.  The challenge would be the detection of the 1529~nm photons to distinguish from the 780~nm cooling beams. The advantages would be that the oscillator strength of this transition for a free-space beam has already been measured experimentally \cite{NILSEN1978327} and a similar transition in Cs, i.e., the $6S_{1/2} \rightarrow 5D_{5/2}$ transition, has been studied in the evanescent field of a prism \cite{tojo_absorption_2004}.   This work extends the use of optical nanofibres in atomic systems and could find applications in atomic clocks, lifetime measurements of atomic states, and in devising trapping schemes for neutral atoms.   

\section{Acknowledgements}
This work was supported by Okinawa Institute of Science and Technology Graduate University and JSPS Grant-in-Aid for Scientific Research (C) Grant Number 19K05316.  The authors wish to acknowledge F. Le Kien for useful discussions, M. Ozer and K. Karlsson for technical support, and E. Nakamura for research support.  

%Use the following for bibtex
\section*{References}
\bibliographystyle{iopart-num}
\bibliography{quadrupole}
%%

%Use the following with biblatex
%\printbibliography

\end{document}